\documentclass[aps,prb,twocolumn,floats,showpacs,superscriptaddress]{revtex4-1}
\usepackage{graphicx,epsfig,psfrag}
\usepackage{times}
\usepackage{graphics,dcolumn,bm,epic,eepic,float}
\usepackage{amssymb,amsmath,rotate,color}
\usepackage{graphicx}

\begin{document}
\unitlength 1 cm
\newcommand{\be}{\begin{equation}}
\newcommand{\ee}{\end{equation}}
\newcommand{\bearr}{\begin{eqnarray}}
\newcommand{\eearr}{\end{eqnarray}}
\newcommand{\nn}{\nonumber}
\newcommand{\la}{\langle}
\newcommand{\ra}{\rangle}
\newcommand{\cd}{c^\dagger}
\newcommand{\vd}{v^\dagger}
\newcommand{\ad}{a^\dagger}
\newcommand{\bd}{b^\dagger}
\newcommand{\fd}{f^\dagger}
\newcommand{\tk}{{\tilde{k}}}
\newcommand{\tp}{{\tilde{p}}}
\newcommand{\tq}{{\tilde{q}}}
\newcommand{\eps}{\varepsilon}
\newcommand{\vw}{{\vec W}}
\newcommand{\vv}{{\vec v}}
\newcommand{\vk}{{\vec k}}
\newcommand{\vp}{{\vec p}}
\newcommand{\vq}{{\vec q}}
\newcommand{\vkp}{\vec {k'}}
\newcommand{\vpp}{\vec {p'}}
\newcommand{\vqp}{\vec {q'}}
\newcommand{\bk}{{\bf k}}
\newcommand{\bp}{{\bf p}}
\newcommand{\bq}{{\bf q}}
\newcommand{\br}{{\bf r}}
\newcommand{\bR}{{\bf R}}
\newcommand{\up}{\uparrow}
\newcommand{\down}{\downarrow}
\newcommand{\fns}{\footnotesize}
\newcommand{\ns}{\normalsize}
\newcommand{\cdag}{c^{\dagger}}
\newcommand{\lc}{\langle\!\langle}
\newcommand{\rc}{\rangle\!\rangle}

\title{Localized magnetic states in three dimensional Dirac solids}
\author{M. Mashkoori}
\affiliation{Department of Physics, Sharif University of Technology, Tehran 11155-9161, Iran}
\author{I. Mahyaeh}
\affiliation{Department of Physics, Sharif University of Technology, Tehran 11155-9161, Iran}
\author{S. A. Jafari}
\affiliation{Department of Physics, Sharif University of Technology, Tehran 11155-9161, Iran}
\affiliation{Center of excellence for Complex Systems and Condensed Matter (CSCM), Sharif University of Technology, Tehran 1458889694, Iran}

\begin{abstract}
   Formation of localized magnetic states in a metallic host is a classic problem of
condensed matter physics formalized by P. W. Anderson within the so called 
single impurity Anderson model (SIAM). 
The general picture in a host of a simple one-band metal
is that a large Hubbard $U$ in the impurity orbital is pre-requisite for the 
formation of localized magnetic states. In recent years three dimensional (3D) Dirac
solids have emerged the hallmark of which is strong spin-orbit interaction. In this
work we show that such a strong spin-orbit interaction allows to form localized magnetic
states even with small values of Hubbard $U$. This opens up the fascinating possibility
of forming magnetic states with $s$ or $p$ orbital impurities -- a different from
traditional paradigms of $d$ or $f$ orbital based magnetic moments. 
\end{abstract}

\maketitle

\section{Introduction}
When an element possessing usually $d$ or $f$ orbital
is added as an impurity to a metal, under certain conditions the
impurity atom can have a magnetic moment. This problem was formulated and
solved within a mean field approximation by Anderson~\cite{anderson61}.
The essential ingredient was identified to be the Hubbard $U$ which is
relatively large in $d$ and $f$ orbital impurities. Anderson's formulation
lead to the following picture: If the Hubbard $U$ is large enough
double-occupancy and empty charge configurations of the impurity 
become energetically costly. If the hybridization $V$ with
continuum of Bloch states in the host metal is weak enough to prevent
decay of the localized spin-split states into continuum, within
the Hartree mean field it leads to the formation of localized 
magnetic states in metallic hosts. In a simple metal considered
in the original Anderson impurity model the spin-orbit is absent
and therefore the study of interplay between the spin-orbit interaction
and other parameters of the Anderson impurity problem remains an 
outstanding problem. 

One interesting paradigm where strong spin-orbit interaction manifests
itself in a fascinating way is concerned with three dimensional Dirac solids. 
Dirac electrons in solids appear under quite general conditions where in 
presence of strong spin-orbit interactions, two bands of Kramers doublets are 
separated by a small gap~\cite{fuseyaReview}.
Under such general conditions the effective bands of the solid can be 
represented by the Wolff Hamiltonian:
\be
H_{\rm W}=\Delta\gamma^0+  \vk.\sum_{j=1}^3 \vv_j\gamma^0\gamma^j
\label{wolff.eqn}
\ee
where $\vv$ is related to velocity matrix elements and $\gamma^j$ and $\gamma^0$ 
are $4\times 4$ matrices  given by~\cite{wolff1964}
\be
   \gamma^0=\left(\begin{array}{cc} 
      I & 0\\
      0 & -I
   \end{array}\right),~~~~~~
   \gamma^j=i\left(\begin{array}{cc} 
      0 & \sigma^j\\
      \sigma^j & 0
   \end{array}\right)
\ee
with $I$ and $\sigma^j, j=1,2,3$ being unit and Pauli $2\times 2$ matrices.  
The four $\gamma^\mu$ matrices with $\mu=0,1,2,3$ satisfy the algebra
of Dirac matrices, namely:
\be
   \gamma^\mu\gamma^\nu+\gamma^\nu\gamma^\mu=2\eta^{\mu\nu}
\ee
where the matrix representation of the tensor in the right side
is $\eta^{\mu\nu}={\rm diag}(1,-1,-1,-1)$. Note that this representation
of Dirac matrices is slightly different from those used in the field
theory texts~\cite{peskin1995}.

The isotropic Wolff Hamiltonian corresponds to situation where velocity matrices
are the same for three Cartesian directions: 
$\vv_j=v \hat e_j$ with $\hat e_j$ being three mutually orthogonal 
unit vectors along the $x,y,z$ directions. Under isotropic conditions 
the Wolff Hamiltonian becomes,
\be
H_{\rm D}=\left(\begin{array}{cc}
\Delta & i v \vk. \textbf{$\sigma$}\\
- i v \vk. \textbf{$\sigma$} & - \Delta
\end{array}\right).
\label{dirac.eqn}
\ee
This is precisely the Dirac Hamiltonian where the
light velocity is replaced by $v$. The above
effective Hamiltonian is obtained by $\vk.\vp$ expansion
around a particular wave-vector $\vk_0$ (corresponding to L point
in bismuth). 

One of the long standing puzzles of such Dirac systems in the context 
of bismuth has been the Diamagnetic response that was markedly different
from normal metallic states. Unlike the one-band situation 
where a Landau-Peierls formula describes the diamagnetic response 
of the solid at hand, in the case of Dirac systems the inter-band 
effects play a crucial role~\cite{fuseyaPRL}. 
In this context an interesting question
can be formulated with respect to the behavior of impurity states
in a host of Dirac electrons:
What is the role played by the presence of the other band of
Kramers doublets? Another interesting aspect of local moment formation
in 3D Dirac solids is related to the role of spin-orbit interactions that
determines the velocity scale $v$ in this Hamiltonian. What is the
interplay between the spin-orbit interaction encoded in $v$ and the 
formation of local magnetic states? In this work we will show that,
unlike normal metals where basically a strong Hubbard $U$ at the
impurity orbital causes local magnetic moment formation by 
excluding double occupancy, in the case of 3D Dirac solids the 
spin-orbit interaction facilitates the formation of localized magnetic
states even with very small values of Hubbard parameter $U$. 
This has far reaching consequences: In addition to impurity atoms 
with $d$ or $f$ orbital, even systems with $s$ or $p$ 
impurity orbitals having smaller values of Hubbard $U$ have a chance
of forming magnetic moments in 3D Dirac solids -- an opportunity not
available in a host of normal metal. Moreover the presence of two bands in
a 3D Dirac solid produces an additional region in the phase diagram which is
which has no counterpart when the host is a normal metal or even a 2D Dirac 
system (graphene). We clarify that this new portion of phase diagram can be 
considered as the signature of a second band of Kramers doublets.

The paper is organized as follows:
In section II we formulate the single impurity Anderson model (SIAM) in a
host of 3D Dirac solid. We set up mean filed equations parameterizing
hybridization of impurity orbital with local orbitals of the 2D Dirac host
from a purely A-sublattice type to B-sublattice type. In section III we report our numerical 
results leading to section IV on applications to realistic Dirac materials.
in Section V we discuss possible deviations from a simple Dirac Hamiltonian
such as the presence of anisotropy as in the original Wolff Hamiltonian 
or addition of a quadratic term that could possibly give rise to a non-trivial
topology of host.

\section{Formulation of the problem}
The isotropic Wolff Hamiltonian for a general 3D Dirac material is
given by~\cite{wolff1964}:
\begin{equation}
H_0 = \sum _\vk 
\Psi^\dagger_{\vk} \left( 
\begin{array}{cc}
\Delta & i v \vk. \vec\sigma\\
- i v \vk. \vec\sigma & - \Delta
\end{array}
\right)
\Psi_{\vk}.
\label{bishamil.eq}
\end{equation}
with a four component spinor $\Psi^\dagger_{\vk} = \left( c^\dagger_{\vk \uparrow}, 
c^\dagger_{\vk \downarrow}, d^\dagger_{\vk \uparrow}, d^\dagger_{\vk \downarrow} \right) $.
In the above basis the operator $c_{k\sigma}^\dagger(d_{k\sigma}^\dagger)$ is creation operator 
in the upper (lower) band. 
These operators at $\vk=\vk_0$ (corresponding to L point in bismuth) can be thought of 
as anti-bonding (bonding) molecular orbitals composed of two atomic orbitals at two locations 
of a unit cell. This Hamiltonian represents a gapped two bands model where $2\Delta$ is the energy 
gap and $v$ is the velocity of Dirac fermions which is usually 2-3 orders of magnitude smaller than 
the light velocity. The spectrum of this Hamiltonian is : 
\be
\eps_k = \pm \sqrt{v^2 k^2 + \Delta^2}. 
\ee

As the canonical model for the study of magnetic moment formation in the host of
itinerant electrons, we consider the SIAM as follows:
\begin{equation}
H = H_0 + H_{\rm imp} + H_{\rm hyb},
\end{equation}
where impurity's contribution is:
\begin{equation}
H_{\rm imp} = \sum _{\sigma} \epsilon _d f^{\dagger} _{\sigma} f_{\sigma} + U 
n_{\uparrow}n_{\downarrow}.
\end{equation}
Here $f^\dagger$ represents creation operator in impurity level of energy $\eps_0$.
and $U$ is the Hubbard repulsive interaction in impurity site.

To construct the hybridization part, consider the elemental bismuth corresponding
to the Hamiltonian \eqref{bishamil.eq}. Bi has a rhombohedral lattice with bases 
composed of two atoms. The creation operator $d^\dagger(c^\dagger)$ creates states 
in symmetric (anti-symmetric) orbitals corresponding to the top of valence band (bottom
of conduction band). Therefore if we define $a^\dagger (b^\dagger)$ as creation operator on 
sub-lattice $A (B)$ then $d^\dagger \sim a^\dagger + b^\dagger$ and $c^\dagger \sim a^\dagger - b^\dagger$.
Therefore a local hybridization with an orbital on site $A$ or $B$ is given as
$f^\dagger_\sigma(c_{\vk\sigma}\pm d_{\vk\sigma})$. Here we assume a quite
general combination of $c$ and $d$ states as $\lambda c+\xi d$. Therefore the hybridization of 
impurity with the host electrons is assumed to be:  
\begin{equation}
H_{\rm hyb} = \dfrac{1}{\sqrt{\cal{N}}} \sum _\vk [V (\lambda^* c^{\dagger}_{\vk \sigma} 
+ \xi^* d^{\dagger}_{\vk\sigma})f_{\sigma} +V^* f^{\dagger}_{\sigma}(\lambda c_{\vk \sigma} 
+ \xi d_{\vk \sigma} ) ],
\end{equation}
where $V$ is hybridization strength between impurity level and the Bloch states.
As an example one can set $\lambda = +1(-1),\xi = +1$ to 
hybridize the impurity with just one atom from sub-lattice $A(B)$. On the other hand $\lambda=+1,\xi=0$
represents the anti-symmetric hybridization with two sub-lattices (conduction), 
while $\lambda=0,\xi= +1$ stands for symmetric hybridization with two sub-lattices (valence). 
If one writes the equation of motion for the impurity's Green function in frequency domain, 
$\langle{\langle{{f_\sigma }|f_{\sigma'}^\dag}\rangle}\rangle)$ it gives:
\begin{multline}
 (\omega  - \eps_0 )\langle{\langle{{f_\sigma }|f_{\sigma'}^\dag}\rangle}\rangle  = 
 \delta_{\sigma\sigma'}+ \\ U\langle{\langle{{f_\sigma}{n_{\bar\sigma}}|f_{\sigma'}^\dag}\rangle}\rangle + 
 \sum_{\vk} \frac{V^*}{\cal \sqrt{N}} 
 \la\la e_{\vk \sigma}|\fd_{\sigma'} \ra\ra,
 \label{eqom.eq}
\end{multline}
where we define $e_{\vk \sigma} \equiv \lambda c_{\vk \sigma} +\xi d_{\vk \sigma} $.
The Hartree approximation in this case corresponds to replacement 
$\eps_0\to\eps_\sigma=\eps_0+U \langle n_{\bar\sigma}\rangle$. 
Writing the equation of motion for $\la\la e_{\vk \sigma}|\fd_{ \sigma'} \ra\ra $ yields:
\begin{multline}
	\la\la e_{\vk \sigma}|\fd_{\sigma'} \ra\ra =\frac{V}{\cal \sqrt{N}} \frac{1}{\omega^2 
	-\epsilon^2_{k}} \\
	\Big\lbrace \la\la f_{ \sigma}|\fd_{ \sigma'} \ra\ra
	  [ (\omega -\Delta)|\lambda|^2  +(\omega + \Delta)|\xi|^2 
	 + i\sigma (\lambda\xi^*-\lambda^*\xi)v k_z] \\
	  +i  \la\la f_{ \bar{\sigma}}|\fd_{ \sigma'} \ra\ra (\lambda\xi^*-\lambda^*\xi) v 
	  (k_x-i\sigma k_y)  \Big\rbrace.
	  \label{eqomb.eq}
\end{multline}
By combining Eqns.~\eqref{eqom.eq} and \eqref{eqomb.eq}, the impurity's Green function is given by:
\begin{equation}
 (\omega  - \epsilon_\sigma - \Sigma_{f}(\omega))\langle{\langle{{f_\sigma }|
 f_{\sigma'}^\dag}\rangle}\rangle  = \delta_{\sigma\sigma'},
 \label{eqm.eq}
\end{equation}
where the self-energy is given by:
\be
\Sigma_{f}(\omega) = \frac{|V|^2}{\cal{N}} \sum_{\vk} \frac{\omega(|\lambda|^2 + |\xi|^2) + \Delta (|\xi|^2 - |\lambda|^2)}{\omega^2 -\eps^2_{k}} .
\label{selfn.eq}
\ee

\begin{figure}[t]
\center
\includegraphics[width = 10.0cm]{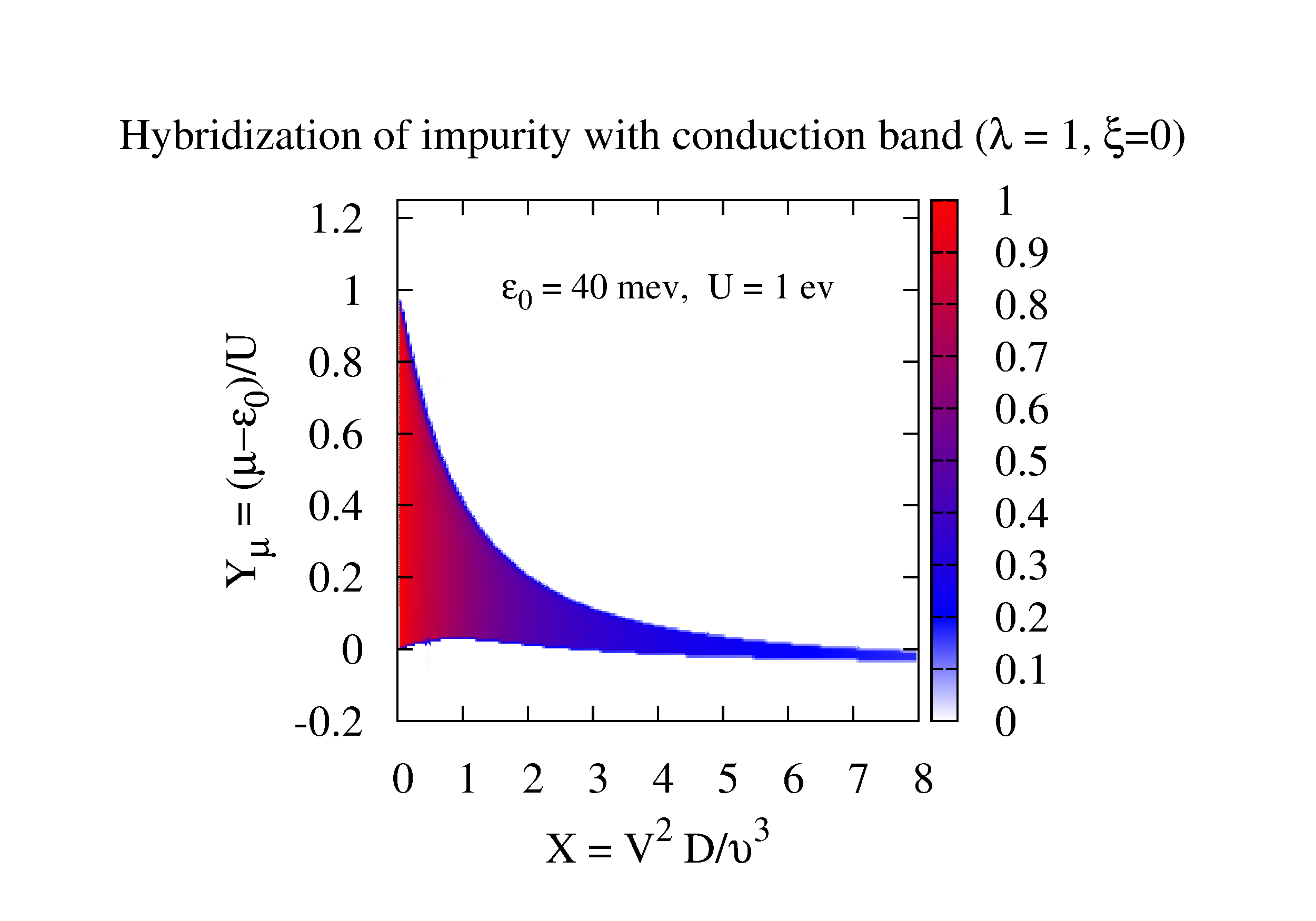}
\caption{(Color online) Phase diagram for local moment formation in 3D Dirac solids.
The area enclosed by the curve and $Y$-axis is magnetic region. The subscript $\mu$ in $Y_\mu$
emphasizes that for obtaining this diagram chemical potential has been tuned. 
In this case, impurity level lies in conduction band.}
\label{pd1.fig}
\end{figure}

In order to obtain the above self-energy the fact that if a given state at $\vk$ is occupied, 
the time-reversed state at $-\vk$ is occupied too simplifies the integration.
Summation over $\vk$ for obtaining self-energy leads to diagonality with respect to spin and 
simplifications in the Green function for general hybridization pattern parameterized by
arbitrary $\lambda$ and $\xi$. Performing the integration over $\vk$ in Eq.~\eqref{selfn.eq}, 
the self-energy as a function of $\omega$ becomes:
\begin{multline}
\Sigma_f(\omega) = \dfrac{\omega(|\lambda|^2 + |\xi|^2) + \Delta (|\xi|^2 - |\lambda|^2)}{2\pi \omega D}
{\tilde{v}\sqrt{\omega^2 -\Delta^2}} ~\times\\
\Big[2\omega \ln{(\frac{{\sqrt{D^2-\Delta^2}}+{\sqrt{\omega^2-\Delta^2}}}
{{\sqrt{D^2-\Delta^2}}-{\sqrt{\omega^2-\Delta^2}}})} 
 -4\omega \frac{\sqrt{D^2-\Delta^2}}{\sqrt{\omega^2 -\Delta^2}}   - i \pi \vert \omega \vert \Big].
\label{self.eq}
\end{multline}

In the above equation $D$ is the bandwidth cut-off and the parameter $\tilde{v} \equiv V^2D/(2\pi v^3)$
naturally emerges in the theory that in addition to the Fermi golden rule decay rate proportional
to hybridization strength $V$ contains the velocity scale $v$ (that arises from the spin-orbit
coupling). This means that in a host of 3D Dirac electron the hybridization strength 
and spin-orbit coupling do not independently determine the physics of local moment formation; 
instead the specific combination $ \sim V^2/v^3$ plays the role played by the combination 
$V^2/U$ in normal metals. {\em This is already very suggestive that in 
3D Dirac solids a large spin-orbit coupling may lead to local magnetic moments 
in a similar way the Hubbard $U$ in normal metals does}. As will be seen 
in the next section, this is indeed the case, and unlike normal metals, the
3D Dirac solids allow for magnetic moment formation even for impurity orbitals
with small values of Hubbard $U$. 
We therefore use the combination $X=V^2D/v^2$ to construct the phase diagram of
local magnetic moment formation in 3D Dirac materials. In order to contrast the 
phase diagram against normal metallic hosts, we choose $Y=(\mu-\eps_0)/U$.
Moreover, Eq.~\eqref{self.eq} shows that the relative phase of complex numbers $\lambda$ and $\xi$ 
is not important in phase diagram which is determined by self-energy. We compute occupation number 
for both spins in impurity's site, i.e. $n_{\uparrow}$ and $n_{\downarrow}$. 
Occupation number can be calculated as follows:
\be
\label{occu}
\la n_{\sigma}\ra = -\frac{1}{\pi} \int_{-\infty}^{\mu} d\omega \frac{\Im\Sigma_{f}}
{\left[Z^{-1}(\omega)\omega - \eps_{\bar{\sigma}}\right]^2 + \Im(\Sigma_{f})^2 },
\ee
where,
\begin{multline}
Z^{-1}(\omega) = 1 -  \dfrac{\omega(|\lambda|^2 + |\xi|^2) + \Delta (|\xi|^2 - |\lambda|^2)}{2\pi \omega D}
{\sqrt{\omega^2 -\Delta^2}} \times \\ 
\tilde{v} \left[2 \ln{(\frac{{\sqrt{D^2-\Delta^2}}+{\sqrt{\omega^2-\Delta^2}}}
{{\sqrt{D^2-\Delta^2}}-{\sqrt{\omega^2-\Delta^2}}})}-4 \frac{\sqrt{D^2-\Delta^2}}{\sqrt{\omega^2 -\Delta^2}} \right].
\end{multline}
 Difference of occupation number is local magnetic moment, $m=n_{\uparrow} -n_{\downarrow}$.

\section{Numerical results}
We solve Eq.~\eqref{occu} for $\up$ and $\down$ spins self-consistently. 
By tuning $\tilde{v}$ ($X$ axis) and $\mu$ (which results in scanning $Y$ axis) 
we construct the region of parameter space corresponding to localized magnetic
states in three dimensional Dirac materials. We consider the case of impurity level with $\eps_0=40$ meV and 
$U=1$ eV which hybridizes anti-symmetrically with two sub-lattices; this leads to hybridization with conduction band. 
Phase diagram is presented in Fig.\ref{pd1.fig}. This result is suggesting that stronger spin-orbit 
coupling which means larger velocity $v$ of Dirac fermions, leads to larger magnetic moments (red color
in the intensity plot means magnetic moments closer to one -- in units of $\hbar/2$). 
Due to spin-orbit coupling, the true eigen-states of the Hamiltonian $H_0$ of the host
Dirac material is actually a Kramers doublet which is a linear combination of spin $\up$ and $\down$ 
states. For large enough $U$ (small $Y$) where conditions for single-occupancy of the 
impurity orbital are favorable, the mechanism that can reduce the local moment e.g. in the $\up$
state is the tunneling out of the impurity $\up$ state to another $\up$ state in the surrounding
continuum of states of the host material. But since the eigen-states of the $H_0$ are not purely
$\up$ nor purely $\down$ states, the spin-orbit coupling weakens the rate of such transitions
out of and into the localized state. Therefore stronger spin-orbit coupling is expected to 
give rise to stronger localized magnetic moments under comparable conditions. 
The formal way of understanding the above situation is that the broadening 
of spin-split impurity states are given by $\Gamma\sim V^2\rho(\eps)/v^3$. Therefore
larger spin-orbit coupling ($v$) leads to smaller broadening, and hence a more
perfect spin-split levels, i.e. larger magnetic moments. 
This form of spin-orbit coupling appears only in 3D Dirac solids, and hence
the present mechanism of the local moment enhancement can be considered as 
a characteristic of these systems.

The second property of the region of magnetic moments in the above figure is that
it is confined to $Y>0$. While this feature is similar to the behavior of localized
magnetic moments in a host of normal metallic hosts, 
it is distinct from the magnetic moment formation in the two dimensional 
Dirac systems (graphene)~\cite{Uchoa2008}. As pointed out in Ref.~\cite{Uchoa2008}
this difference can be traced back to the damping behavior of the local Green's function
at large $\omega$. This behaves as $\omega^{-2}$ in normal metals and 3D Dirac solids,
while in the 2D Dirac systems it goes as $\omega^{-1}$. 
Despite that local magnetic moment formation in 3D Dirac and normal metals
both occur for $Y>0$ (i.e. $\mu>\eps_0$), it is interesting to note that
the upper boundary of the magnetic region has positive curvature in 3D Dirac solids,
while in the normal metals the curvature is negative and the upper bound is convex. 

\begin{figure}[t]
\center
\includegraphics[width = 10.0cm]{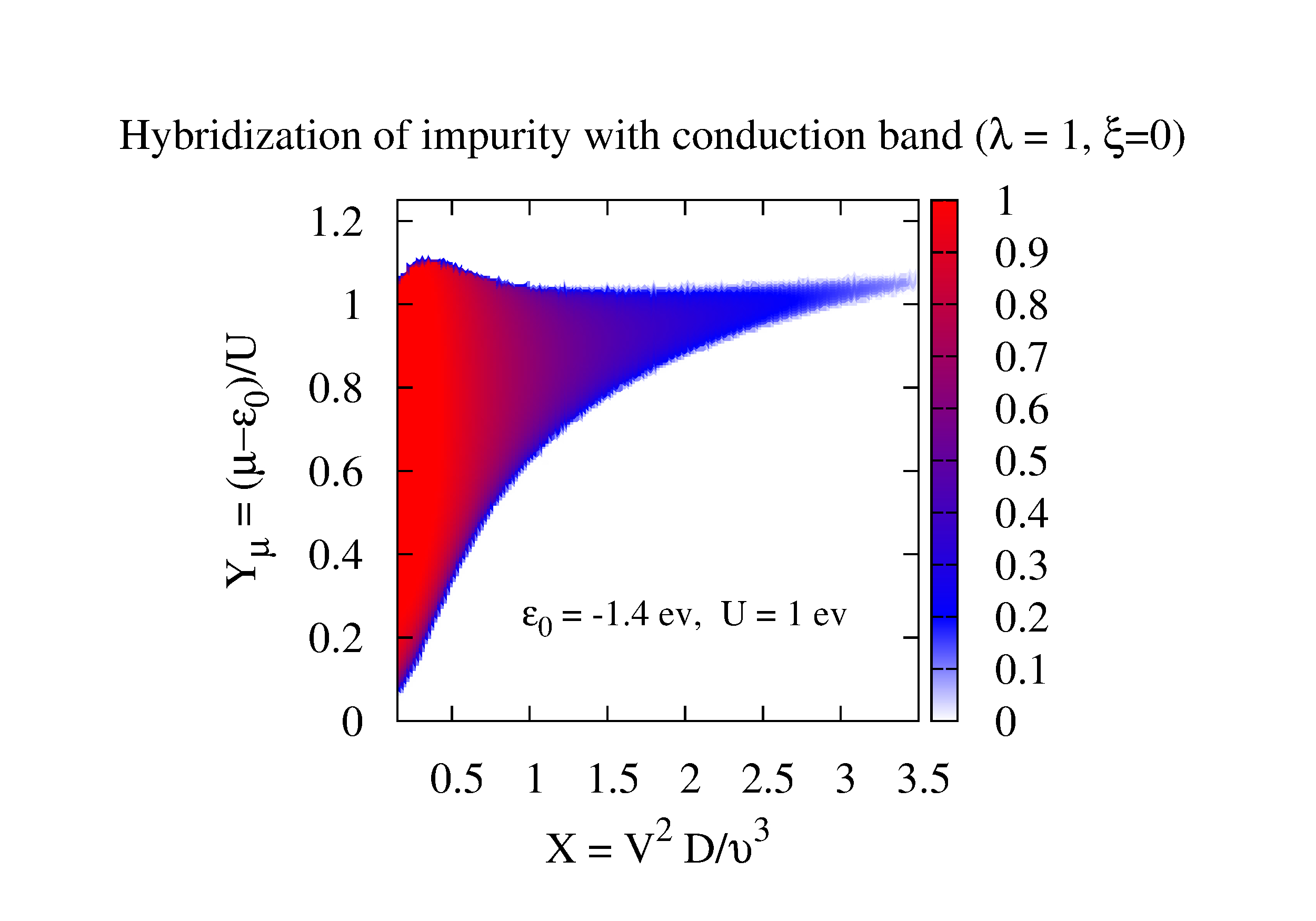}
\caption{(Color online) Phase diagram for local moment formation in 3D Dirac solids. 
The impurity level $\eps_0=-1.4$ eV lies deep in the valence band. 
The color code is indicated in the legend. The subscript $\mu$ in $Y_\mu$
means that to obtain this diagram chemical potential has been tuned. 
In this figure the tail of the magnetic phase is extended along $\mu - \eps_0 \approx U$ line. }
\label{pd2.fig}
\end{figure} 

In Fig.~\ref{pd2.fig} we construct the phase diagram for a very negative value of $\eps_0=-1.4$ eV. 
This figures shares the general property with Fig.~\ref{pd1.fig} that larger spin-orbit
coupling gives rise to stronger magnetic moments. However they differ in their large $X$ behavior.
In Fig.~\ref{pd1.fig} corresponding to $\eps_0=40$ meV, weak magnetic moments for large $X$ values
are formed when $\mu-\eps_0\approx 0^+$, while in Fig.~\ref{pd2.fig}
corresponding to $\eps_0=-1.4$ eV, the corresponding small moment states are formed for 
$\mu-\eps_0\approx U$. The large $X$ is equivalent to small spin-orbit coupling and
large hybridization $V$. 
In the limit of large $X$, the spin-orbit coupling becomes negligible.
Assuming that the Hubbard $U$ is large enough to favor single occupancy, 
the dominant term to be minimized will be hybridization term giving rise
to energy contribution $|V|^2\rho(\mu)$. For a given $\eps_0$ and allowing
$\mu$ to be variable, the minimization of the above energy contribution 
amounts to selecting regions with smaller DOS. For small values of
$\eps_0=40$ meV in the conduction band subject to the $Y>0$ condition,
the smallest value of $\rho(\mu)$ is obtained when $\mu\to \eps_0$.
In the case of very negative $\eps_0=-1.4$ eV, minimization of $\rho(\mu)$
is achieved when $\mu\to\eps_0+ U$.  The sign of $\eps_0$ affects the 
elongation pattern of magnetic region in the same way as graphene~\cite{Uchoa2008}.

\begin{figure}[t]
\center
\includegraphics[width = 8.0cm]{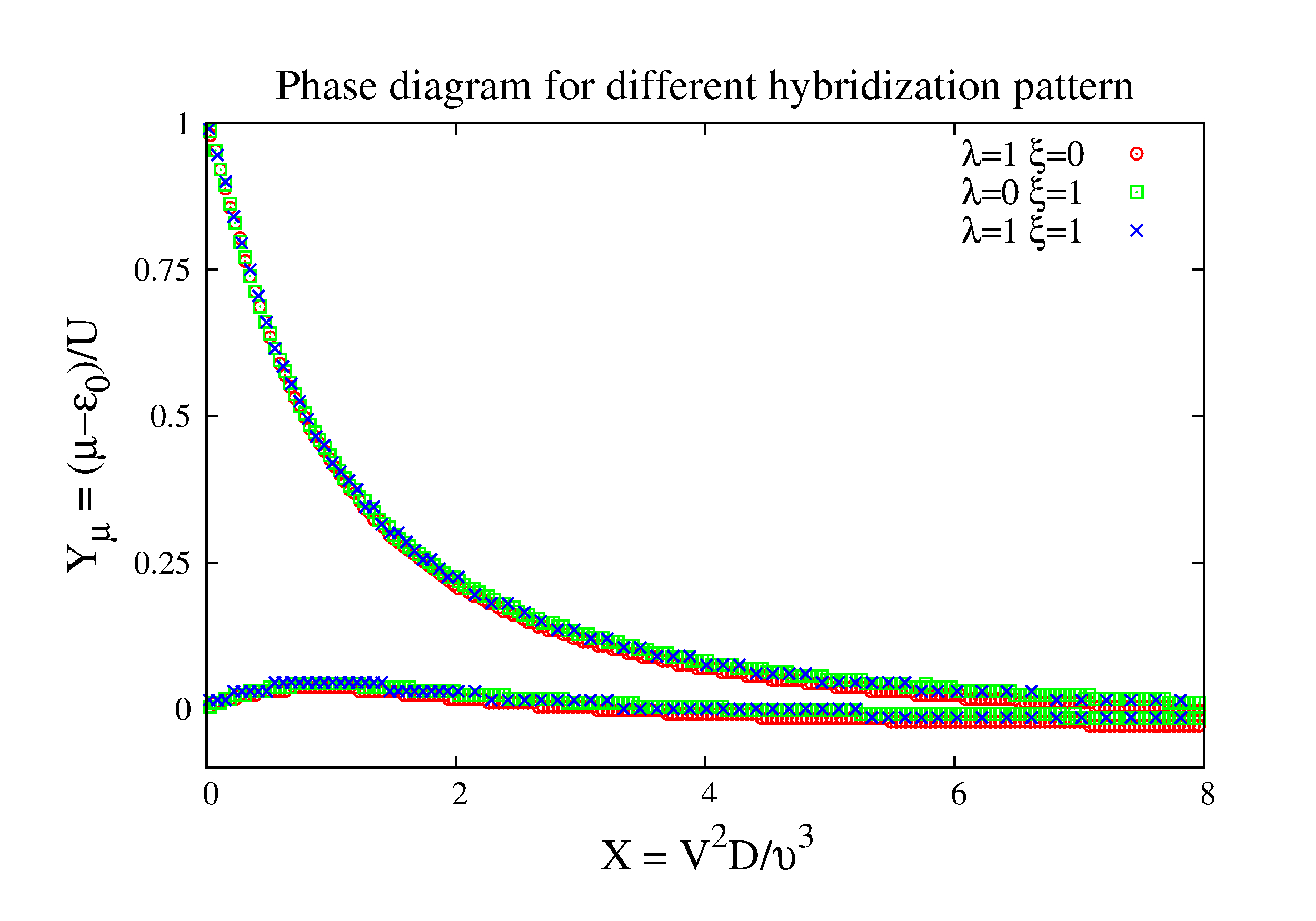}
\caption{(Color online) Comparison of different hybridization in which we normalize strengths. 
In this plot the chemical potential is tuned and the impurity level and Hubbard $U$ set
$\eps_0 = 40$meV and $U=1$eV respectively. Three different possibilities for coefficient $\lambda$ 
and $\xi$ is investigated. The figure demonstrates 
that the phase diagram does not change for local hybridization of impurity with an atom 
for sublattice A or B, or a combination  of them.}  
\label{compare.fig}
\end{figure} 

So far we have been concerned with the situation where the impurity orbital
was coupled to the conduction band, i.e. $\lambda=+1,\xi=0$. It is
interesting to see what happens when the relative weights of the $\lambda$ and $\xi$ changes. 
In Fig.~\ref{compare.fig} we show comparison with different hybridization patterns. We consider three cases: 
(a) $\lambda= 0,\xi= 1$ (b) $\lambda= 1,\xi= 0$ and (c) $\lambda= 1,\xi= 1$. 
The cases (a) and (b) correspond to hybridization of the impurity with 
a molecular orbital of either c or d character, while the case (c) above
corresponds to hybridization with an atomic orbital in sub-lattice A. 
To have a meaningful comparison between the above three cases, we should perform 
an appropriate scaling: 
From Fermi's golden rule, the broadening of the spin-split impurity levels
is proportional to $V^2$. This broadening in (a) and (b) cases is half of the (c) case. 
So we should scale $X$ axis for later case. In doing so, the local moment boundaries 
for all the above three cases collapse on the same curve as depicted in Fig.~\ref{compare.fig}.   
Parameter values are indicated in the figure caption. 
This indicates that as long as hybridization remains local, there is no
conceptual difference between coupling the impurity to an atom from sublattice A, or B
or a combination thereof.

\section{Application to materials and deformations of the Dirac Hamiltonian}
As pointed out in the introduction, the derivation by Wolff of the anisotropic
Dirac Hamiltonian for 3D Dirac solids is quite general and applies to a broad
range of materials the difference of which is reflected in model parameters. 
The initial motivation of Wolff was to construct an effective single-particle theory for 
the low-lying electronic states of bismuth. In this section let us discuss how do 
the material specific considerations affect the results. We obtain phase diagrams for various 
situations corresponding to variations in different parameters of the model, and adopting 
numbers related to bismuth. 
\begin{figure}[t]
\center
\includegraphics[width = 10.0cm]{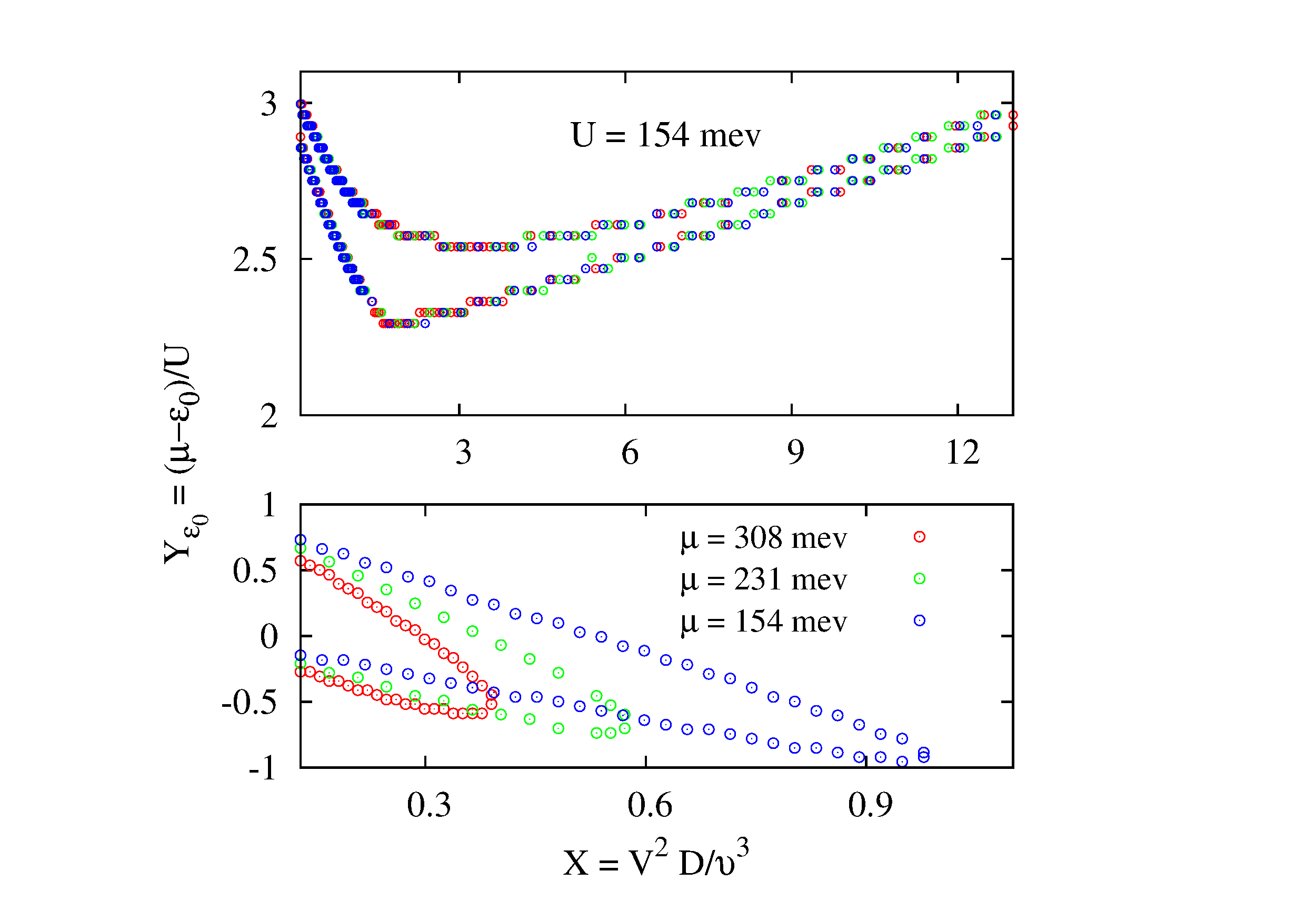}
\caption{(Color online) Phase diagram for the local moment formation in SIAM for bismuth. 
The area enclosed by the curves and the $Y$-axis is magnetic region. In this plot the variables $X$ and 
$Y$ are defined by $X = \vert V\vert ^2 D/(2 \pi v^3)$ , $Y=(\mu-\eps_0)/U$. 
As discussed in the text, the top panel shows the hybridization of impurity level
with valance band the the bottom panel represents the hybridization with conduction band.
}
\label{pd3.fig}
\end{figure}

\subsection{Tuning impurity level}
The first thing in a host of bismuth with gap parameter $\Delta= 7.5$ meV
can be changed is the type of impurity. This obviously changes the value $\eps_0$. 
So let us consider the tuning of the parameter $\eps_0$ of the SIAM.
Relative permeability of bismuth and its compounds is in the range of 10-40 which results
in decrease of Coulomb repulsion up to two orders of magnitude~\cite{Young1973}.  
Following the work of Haldane and Anderson on impurity in semiconductor 
materials~\cite{haldane1976,sato2010}, we expect that reduction of charge 
accumulation on impurity site to give rise to decrease of Coulomb repulsion 
up to two orders of magnitude. Therefore, we investigate the problem for very small 
values of Coulomb interaction. We focus on the case where $U=154$ meV 
( ${{U} \mathord{\left/{\vphantom 
{{U} {\Delta}}} \right. \kern-\nulldelimiterspace} {{\Delta}}} = 20 $)
and present phase diagram for three values of chemical potential, $\mu/\Delta=20,30,40$ in 
Fig.~\ref{pd3.fig}. This phase diagram has been constructed by varying $X$ and $\eps_0$ 
(the variation of $Y$ through variation of $\eps_0$ is emphasized by a subscript: $Y_{\eps_0}$)
for three representative values of the chemical potential $\mu$. 
As can be seen, upon tuning $\eps_0$, the magnetic region splits into
two regions (i) an elbow shaped region for larger $Y$ values depicted in the upper
part of the figure, and (ii) a lobe-shaped region for smaller values of $Y$. 

We have deliberately separated the regions (i) and (ii) above, and have applied a
shift $\delta\mu/U$ of the whole curves along the vertical axis to reveal the
different behavior of the two regions upon such a vertical shift.
Since we have selected the above three values of the chemical potential to 
be in the conduction band, the contribution from integrations over the valence
band will be identical for three chemical potentials. Therefore we expect
different phase boundary curves corresponding to different values of chemical
potential to coincide after a vertical shift that compensates the difference
in the chemical potentials. The fact that after such a shift the upper part of the
phase diagram coincide indicates that this region is basically formed by 
continuum of states in the valence band. 
On the other hand, the lower lobe-shaped region for three different chemical potentials
do not coincide after a simple shift and hence they are contributed by the 
conduction band states. This piece of phase diagram is qualitatively close
to the magnetic region of one-band hosts and 2D graphene. Therefore the additional
elbow-shaped part of the phase diagram can be considered a feature of two-band 3D Dirac 
systems. Having separate contributions to the local moment formation from the two
bands of the host material is reminiscent of the qualitatively different
diamagnetic behavior of bismuth compared with normal (one-band, non-Dirac) metals
which can be understood based on a two-band picture and the 3D Dirac Hamiltonian~\cite{fuseyaReview}.
The presence of the other band leads to two different regions (i) and (ii) in the
local moment phase diagram of bismuth. This feature is qualitatively different from
that of normal metals. The lobe-shaped feature of the region (ii) is qualitatively similar
to the magnetic moment region of 2D Dirac fermions in graphene, but the 
elbow-shaped remains a feature peculiar to 3D Dirac fermions. 

\begin{figure}[b]
\center
\includegraphics[width = 10.0cm]{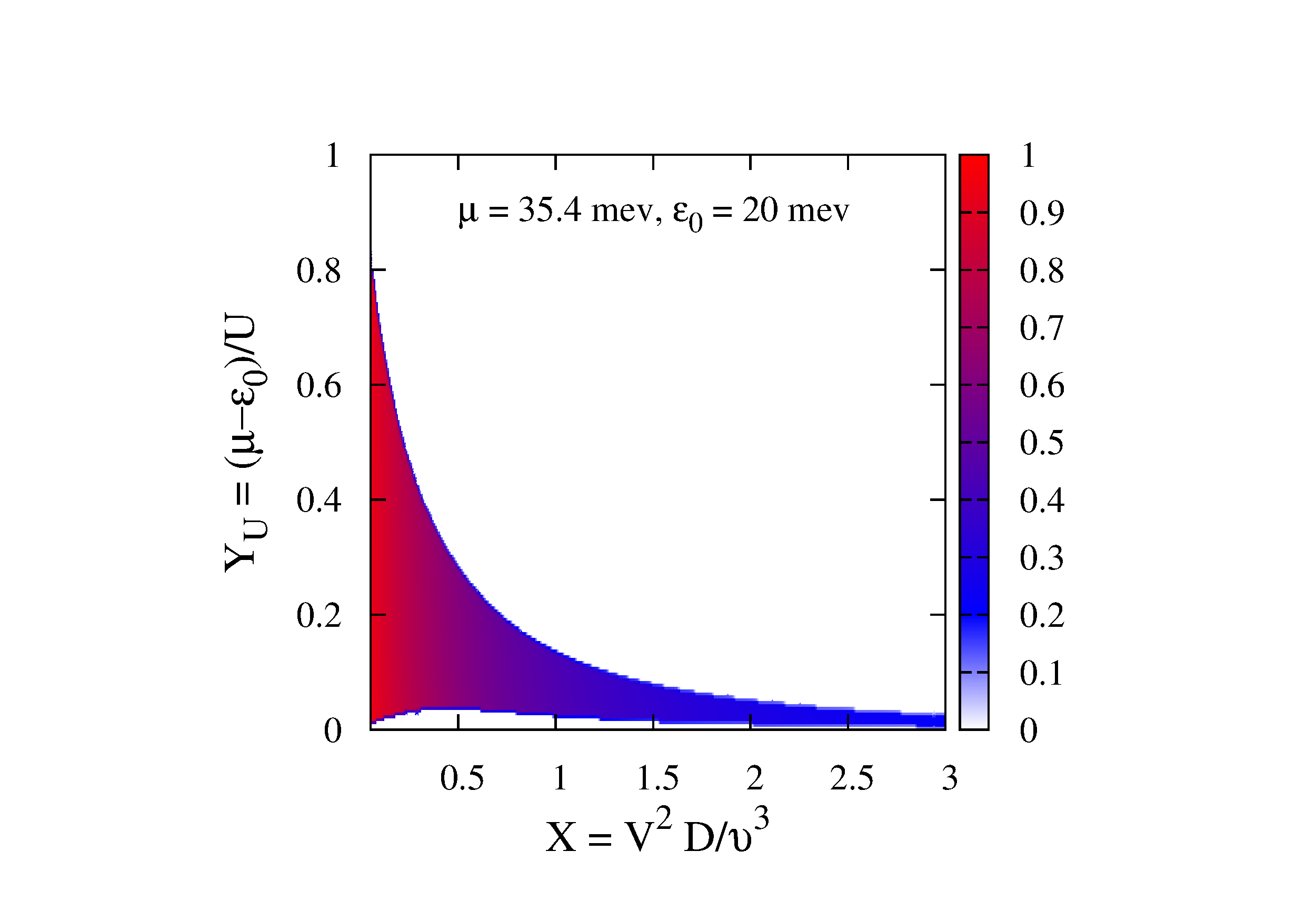}
\caption{(Color online) Local magnetic moment phase diagram obtained by 
tuning the Hubbard $U$.
In this diagram the chemical potential and impurity level are set to be
$35.4$ meV and $20$ meV, respectively. By tuning the Coulomb repulsion $U$ in
this case where $\mu-\eps_0>0$, the magnetic region is qualitatively similar to 
Fig.~\ref{pd1.fig} obtained by tuning $\mu$.}
\label{pd4.fig}
\end{figure}

\subsection{Tuning $U$}
In the single impurity Anderson model for normal metals, the combination $Y=(\mu-\eps_0)/U$
naturally appears in the mean field theory. Therefore as long as the variable $Y$ varies,
it does not matter which of the three parameters $\mu,\eps_0,U$ is responsible for variation
of parameter $Y$. But since in the case of three dimensional Dirac materials the parameter
$Y$ does not emerge naturally, when constructing the traditional phase diagrams in the
$XY$ plane the quantity that gives rise to variation in $Y$ becomes important. 
For 3D Dirac solids one natural parameter is $V^2/v^3$, but the other parameters
can in principle be varied independently leading to a multi-dimensional phase 
diagram. Therefore keeping some fixed, and varying others corresponds to 
viewing a projection of multi-dimensional phase. This can be viewed as an advantage
as it may reveal new features as we will show in this section.
Let us see what happens when we construct the phase diagram in the $XY$ plane
by tuning the Hubbard parameter $U$.  We consider two cases, $\mu-\eps_0>0$ 
and $\mu-\eps_0< 0$. The first case is shown in Fig.\ref{pd4.fig}. 
Its general features are similar to Fig.~\ref{pd1.fig} obtained by tuning 
the chemical potential $\mu$.
\begin{figure}[t]
\center
\includegraphics[width = 10.0cm]{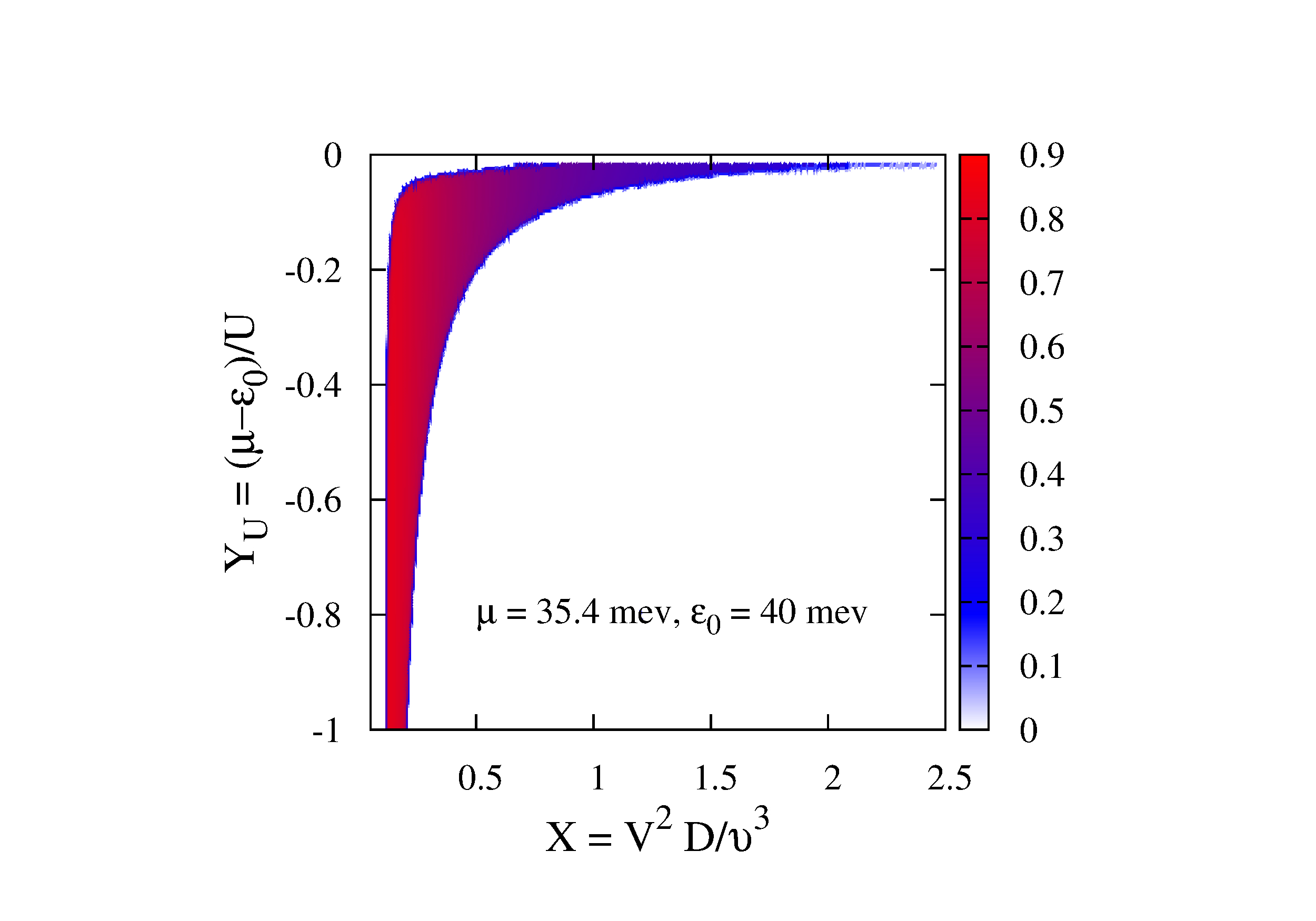}
\caption{(Color online) Local magnetic moment phase diagram obtained by 
tuning the Hubbard $U$.
In this diagram the impurity level lies above the chemical potential $\mu - \eps_0 <0$.
This diagram suggests that in this case for Coulomb repulsion as weak as 5 meV,
 with the aid of strong spin orbit coupling, we may find nonzero magnetization.}
\label{pd5.fig}
\end{figure} 

 However when we repeat the same analysis for $\mu-\eps_0<0$ (and $U>0$ 
which means negative values of $Y$) the phase diagram will be completely
different and brings about a very peculiar physics of local moment formation
in 3D Dirac materials. As can be seen in Fig~\ref{pd5.fig}, for a very 
large values of spin-orbit coupling (reflected in $v$)
even with small values of $U$ quite strong local magnetic moments can form. 
This corresponds to the red vertical part of the phase diagram in 
Fig.~\ref{pd5.fig}. On the contrary looking at the right end of the 
magnetic region in this figure at $X\sim 2$ axis which corresponds to small spin-orbit regime
indicates that in this region there are no local moment for very small spin-orbit couplings even with
large Coulomb repulsions. By increasing spin-orbit coupling (reducing $X$), we come to a region 
in which local moment formation is confined between an upper and lower boundary values for the
Hubbard repulsion. The existence of a minimum repulsion to have unequal populations 
$n_\uparrow$ and $n_\downarrow$ is understandable from the atomic limit. 
If we increase spin-orbit coupling further, as already pointed out, 
we have a region in which a small amount of Coulomb repulsion (as small as $5$ meV) 
leads to formation of 
local moment. This is probably one of the interesting aspects of the local moment
formation in a three dimensional Dirac material with strong spin-orbit interaction
that has no counterpart in normal metallic hosts without spin-orbit interactions.
This means that in a 3D Dirac solid even impurity orbitals such as $s$ or $p$ orbitals
possessing smaller values of Hubbard $U$ as compared to $d$ or $f$ electron adatoms
get a chance of magnetization!
Finally, let us focus on the white region adjacent and parallel to $Y$ axis in 
Fig.~\ref{pd5.fig} that corresponds to the infinitely large spin-orbit coupling,
$v\rightarrow \infty$. As the phase diagram shows in this situation, 
irrespective of the value of $U$, no local moments are formed.

\section{Discussion and summary}
Let us first summarize our findings so far:
We have investigated the formation of localized magnetic states in 
three dimensional Dirac solids, and have found that the
spin-orbit coupling significantly helps with the formation of local
magnetic moments. 
Our investigation shows that the effect of spin-orbit coupling is to enhance
the local moments once they are formed. It also allows for formation of local 
moments with very small values of Hubbard $U$ for strong spin-orbit couplings. 
This means that even $s$ or $p$ orbital adatoms whose Hubbard $U$ is usually
smaller than $d$ or $f$ orbital atoms may find a chance of getting magnetized
in a host of 3D Dirac solid. This chance is not present for them in normal
metals. Construction of phase diagram by tuning the impurity level $\eps_0$ gave rise
to two disjoint pieces of magnetic regions. The elbow-shaped region having no
counterparts in normal metals nor in 2D Dirac system (graphene) is due to 
presence of a second band, while the lobe shaped part of the phase diagram
comes from the band crossing the Fermi level. 
In the SIAM for a host of three dimensional Dirac material, unlink the normal metallic
hosts, the parameter $Y=(\mu-\eps_0)/U$ does not naturally emerge. Therefore in principle
the phase diagram should be constructed in a multi-dimensional parameter space. 
Insisting to represent the phase diagrams in terms of traditional $XY$ parameters
brings in interesting aspects of the localized magnetic states in three dimensional
Dirac solids. Such a larger phase space may provide opportunities for new
applications and directions in the magnetic properties of of Dirac solids.

Now let us speculate on some deformations of the isotropic Wolff Hamiltonian
First thing to discuss is the
role of anisotropy which becomes relevant when it comes to real materials:
In presence of anisotropy, the velocity will be different
for three different directions, and hence the dispersion 
relation of the host Dirac material becomes,
\be
\eps_k = \pm \sqrt{v^2_x k^2_x+v^2_y k^2_y+v^2_z k^2_z+\Delta^2}.    
\ee
If we use this dispersion in Eq.~\eqref{selfn.eq} when it comes to integration over 
$\vk$, we can 
rescale variables as $v_x k_x \rightarrow v \tilde{k}_x$. This leads to a Jacobian 
of the form $v^3/(v_x v_y v_z)$ multiplying the same integral as the one in the isotropic case.
Therefore the self-energy rescaled by Jacobian which can be taken in to account by 
appropriately redefining $\tilde{v}$. Therefore, the role of anisotropy is just a 
matter of scaling $X$ axis and does not affect the qualitative physics discussed in this paper. 

There is another conceptually important deformation of the simple
Dirac Hamiltonian. 
The Dirac Hamiltonian \eqref{dirac.eqn} can also be generalized 
by adding a quadratic term 
\be
   H_{\rm G}=v\sum_{j=1}^3k_j\alpha^j
   +(\Delta-B k^2)\beta.
\ee
where $\alpha^j=\gamma^0\gamma^j$ and $\beta=\gamma^0$ are 
usually defined in Dirac equation in terms of Dirac matrices $\gamma^\mu$.
This generalization allows for two possibilities with respect 
to topology of the resulting host material: When $B\Delta <0$ it
is topologically trivial, while if $B\Delta >0$ it has a non-trivial
topology~\cite{ShenBook} with associated boundary states.  
Having gained some insight into the important role of spin-orbit
coupling in the local moment formation in 3D Dirac solids, 
we can briefly address the role of quadratic $B$ term
in the limit of small $B$. In this Taylor expansion of the 
resulting dispersion relation leads to a straightforward renormalization of velocity
i.e. $v^2 \rightarrow v^2 - B\Delta$. 
Therefore, in topologically non-trivial (trivial) case where $B\Delta>0$ ($B\Delta<0$), 
the quadratic term leads to a decrease (increase) in the effective spin-orbit interaction. 
Therefore the topological twist of spins in the momentum space corresponding to 
non-trivial topology are expected to weaken the aspects of local moment physics
of 3D Dirac solids discussed in this paper, while in the topologically trivial 
case, at least within the present perturbative scheme limited to very small $B$,
the spin-orbit driven aspects of local moment physics are expected to get
enhanced upon addition of the quadratic term $B$.

\section{Acknowledgement}
We thank M. Ogata and T. Tohyama for useful discussions.

\bibliography{references}

\end{document}